\documentclass[12pt]{article}
\usepackage{amsmath}
\usepackage{amssymb}
\usepackage[dvips]{graphics,color}
\begin{document}
%\parindent 0 pt
%\journalname{Communications in Mathematical Physics}
%\renewcommand{\baselinestretch}{1.5}

%%%%%%%%%%%%%%%%%%%%%%%%%%%%%%%%%

\def \ep{\epsilon} 
\def \intR {\int_{-\infty}^{+\infty}}  
\def \grad {\nabla}  
\def \ov{\over} 
\def \q  {\quad}  
\def \qq {\qquad}  
\def \bar{\overline} 
\def \beq{ \begin{equation} } 
\def \eeq{\end{equation}} 
\def \r#1{$^{#1}$}
\newtheorem{proposition}{Proposition}
\newtheorem{lemma}{Lemma}                 
\newtheorem{theorem}{Theorem}
\newtheorem{corollary}{Corollary}
\renewcommand{\labelenumi}{(\roman{enumi})}

%%%%%%%%%%%%%%%%%%%%%%%%%%%%%%%%%%%%%%%%%%%%%%%%%%%%%%%%%%%%%%%%%%%%%%%%%%%

\title{\bf Saari's Conjecture for the Collinear $n$-Body Problem}

\author{Florin Diacu\thanks{Pacific Institute for the Mathematical Sciences
and Department of Mathematics and Statistics, University of Victoria, 
P.O.\ Box 3045 STN CSC, Victoria, B.C., Canada, V8W 3P4, E-mail: 
diacu@math.uvic.ca},~ Ernesto P\'erez-Chavela\thanks{Departamento 
de Matem\'aticas, Universidad 
Aut\'onoma Metropolitana-Iztapalapa, Apdo.\ 55534, 
M\'exico, D.F., M\'exico, E-mail: epc@xanum.uam.mx}  ~and   
Manuele Santoprete\thanks{Department of Mathematics, University 
of California, Irvine, 294 Multipurpose Science \& Technology 
Building, Irvine CA, 92697 USA. 
E-mail: msantopr@math.uci.edu}
}

\maketitle
\def \r#1{$^{#1}$}


\begin{abstract}
%------------------------------------------------------------------------------

\noindent In 1970 Don Saari conjectured that the only solutions 
of the Newtonian $n$-body problem that have constant moment of 
inertia are the relative equilibria. We prove this conjecture in
the collinear case for any potential that involves only the mutual
distances. Furthermore, in the case of homogeneous potentials, 
we show that the only collinear and non-zero angular momentum 
solutions are homographic motions with central configurations.   
%------------------------------------------------------------------------------

\end{abstract}

\small\normalsize
%%%%%%%%%%%%%%%%%%%%%%%%%%%%%%%%%%%%%%%%%%%%%%%%%%%%%%%%%%%%%%%%%%%%%%%
\section{ Introduction}
%%%%%%%%%%%%%%%%%%%%%%%%%%%%%%%%%%%%%%%%%%%%%%%%%%%%%%%%%%%%%%%%%%%%%%%

In a series of papers published in the 1970's, Don Saari investigated  
the boundedness of solutions in the Newtonian $n$-body problem. In one 
of the earliest, \cite{Saari}, he proved that if a  solution has constant 
potential for any finite time interval, then it has both constant potential 
and constant moment of inertia at all times. In the same paper he stated 
what is now known as Saari's conjecture: {\it Every solution of the 
Newtonian $n$-body problem that has constant moment of inertia is a
relative equilibrium.}
 
During a recent visit at the University of Victoria, Don told us that 
in the 1970s he had found a simple proof of his conjecture in the 3-body
case. Unfortunately he never published it, lost his notes, and could not remember the details. Though many people tried to solve his conjecture, nobody succeeded yet. In the late 1970s and early 1980s, Julian Palmore published two papers, \cite{Palmore1, Palmore2}, in which he claimed to 
have a proof, but his arguments do not survive a careful scrutiny. 
In October 2002, Chris McCord gave a talk at the Midwest Dynamical Systems 
Conference held at the University of Cincinnati, where he presented a proof in the case of three bodies of equal masses, \cite{McCord}. Also, Rick Moeckel has announced a computer assisted proof for the Newtonian 3-body case. At the time we are writing this article, their papers have not yet 
been published.

Recently the interest in Saari's conjecture has grown even more due  
to the variational proof given by Alain Chenciner and Richard Montgomery 
for the existence of the figure eight solution, numerically discovered
by Cristopher Moore, \cite{Chenciner}, in 1993. Carles Sim\'o studied this orbit numerically, \cite{Simo}, and found out that its moment of inertia is 
very close to being a constant all along the motion. This fact prompted Chenciner to ask for analytical arguments that the figure eight solution 
has no constant moment of inertia.  

In this paper we make some modest steps towards understanding Saari's
conjecture. In particular we prove that the conjecture is true in the
collinear case for any number of bodies and for any potential that
involves only the mutual distances. This generalizes a result obtained
by Pizzetti in 1904, \cite{Pizzetti, Wintner}, for the Newtonian 3-body 
problem. Our result is not only more general but the proof we provide is 
much simpler than the one of the Italian mathematician. Our presentation 
is organized as follows. 

In Section~2 we introduce the equations of motion and several 
definitions. In Section~3 we solve the collinear case, i.e., the one 
in which the bodies are on a line that rotates in the plane of motion.
We first prove that every non-zero angular momentum and collinear
solution of the $n$-body problem given by any potential that
depends only on the mutual distances alone, is homographic.
This implies that if the moment of inertia is constant, the
only solutions of this type are the relative equilibria. 
Moreover, we show that the only non-zero angular momentum and 
collinear solutions of the $n$-body problem given by a homogeneous
potential of degree $\alpha\ne -2$ are the homographic motions
with central configurations. We end our paper with Section~4,
in which we give a geometrical interpretation of Saari's Conjecture 
in the Newtonian case.

%%%%%%%%%%%%%%%%%%%%%%%%%%%%%%%%%%%%%%
\section{Equations of Motion}
%%%%%%%%%%%%%%%%%%%%%%%%%%%%%%%%%%%%%%

Consider the planar motion of $n$ interacting bodies $P_1,..., P_n$.
Let the mass and position (with respect to the center of mass) of the the body $P_i$ be given by $m_i$ and   ${\bf r}_i=(x_i,y_i)$. 
Let $r_{ij}=|{\bf r}_i-{\bf  r}_j|$ be the  distance between the $i$th and the $j$th bodies and denote 
$x=({\bf r}_1,...,{\bf r}_n)\in {\mathbb R}^{2n}$ and $\dot x=(\dot{\bf r}_1,...\dot{\bf r}_n)\in 
{\mathbb R}^{2n}$. 
Take $L=T-U $ as the Lagrangian function of the system under discussion, where $U$ and $T$ are the potential and the kinetic energy. Then the Euler-Lagrange 
equations for the body $P_i$ are
%----------------%
\beq
m_i\ddot{\bf r}_i=-{\partial U \over \partial {\bf r}_i}.
\eeq
%----------------
With  the  scalar product
%---------------%
\beq
\langle x, \dot x \rangle=\sum_{i=1}^n m_i {\bf r}_i \cdot \dot{\bf r}_i,
\eeq
%----------------
and the notation
\beq
\nabla=(m_1^{-1}\nabla_1,\dots,m_n^{-1}\nabla_n),
\eeq
the equations of motion  take the simpler form
%---------------%
\beq
\dot x=y, \qquad \dot y=-\nabla U,
\eeq 
%---------------
where $\nabla_i$ is the $i$th gradient component.

The moment of inertia and the kinetic energy can be written as
%---------------%
\beq
I= \langle x, x \rangle , \qquad T={1\over 2}\langle y,\ y\rangle.
\eeq
%----------------
A {\it central configuration} of the $n$-body problem is a configuration $x\in {\mathbb R}^{2n}$ that satisfies the algebraic equations
%---------------%
\beq
\nabla U(x)=\omega^2 \nabla I
\label{releq0}
\eeq
%----------------
for some constant $\omega$ (see \cite{Meyer,Smale,Wintner} for more details).

A given solution $x=x(t)$ of the $n$-body problem is called {\it homographic} 
if, in the barycentric coordinate system, the configuration of the bodies
is similar to itself when $t$ varies. By this we mean that there exist a scalar 
$\nu=\nu(t)>0$, an orthogonal 2-matrix $\Omega=\Omega(t)$ such that for every 
$i$ and $t$ one has
\beq
{\bf r}_i=\nu(t)\Omega(t){\bf r}_i^0,
\eeq
where ${\bf r}_i$, $\nu$, $\Omega$ correspond to an arbitrary $t$ and ${\bf r}_i^0$ denotes ${\bf r}_i$ at some initial instant $t=t_0$.

There are two limiting types of homographic solutions. One appears when the configuration is dilating without rotation, i.e., $\Omega(t)$ is the identity matrix. Such solutions are called {\it homothetic}. The other shows up when 
the configurations is rotating without dilatation, i.e., $\nu(t)=1$. These particular homographic solutions are characterized by
\beq
{\bf r}_i= \Omega(t){\bf r}_i^0
\eeq
and are called {\it relative equilibria}. In such cases the system rotates 
about the center of mass as a rigid body, the angular velocity is constant 
and the mutual distances do not changes when $t$ varies. The relative 
equilibria are the solutions we are interested in, and they make the object 
of Saari's Conjecture. 

In the following sections we will deal with two types of potentials that are more
general than the Newtonian one: first, when the potential energy is a function of 
the mutual distances only and, second, when the potential is an homogeneous function of degree $\alpha$. In the latter case the Lagrangian function is of the form 

%---------------%
\beq
L= T-U={1 \over 2} \sum_{i=1}^n m_i\dot{\bf r}_i^2+ \beta \sum_{i>j} {m_im_j |{\bf r}_i-{\bf r}_j|^{\alpha}}, 
\label{lagrangian}
\eeq

%---------------

\noindent where $\beta=1$ if $\alpha<0$, whereas $\beta=-1$ if $\alpha>0$.
The corresponding Hamiltonian function is
%---------------%
\beq
H= T+U= \sum_{i=1}^n {{\bf p}_i^2 \over 2m_i}- \beta \sum_{i>j} {m_im_j |{\bf r}_i-{\bf r}_j|^{\alpha}}. 
\label{lagrangian}
\eeq
%---------------
The Lagrangian and Hamiltonian  above describe the Newtonian n-body problem when $\alpha=-1$ and $\beta=G$, where $G$ is the gravitational constant.
%%%%%%%%%%%%%%%%%%%%%%%%%%%%%%%%%%%%%%%%%%%%%%%%%%%%%%%%%%%%%%%%%%%%%%
\section{The Collinear n-Body Problem}
%%%%%%%%%%%%%%%%%%%%%%%%%%%%%%%%%%%%%%%%%%%%%%%%%%%%%%%%%%%%%%%%%%%%%

In this section we will prove Saari's Conjecture in the collinear case
for any potential that depends on the mutual distances alone. By the
collinear case we mean the one in which the bodies are on a line that
rotates in the plane around the center of mass of the bodies.  

Let ${\bf K}$ be  the total angular momentum and ${\bf  K}_i$ the angular momentum of the body $P_i$ about the center of mass. Denote by
${\bf M}_i$ the moment of the forces for the body $P_i$. ${\bf F}_{ij}$ 
is the force acting on the body $P_i$ while interacting with $P_j$. It 
is well known (see \cite{Wintner}), and easy to check, that every collinear solution is planar. Assume that the potential $U$ is function of the mutual distances alone. Note that ${\bf F}_{ij}={\grad_{{\bf r}_{ij}}}U$. In
this setting, we can prove the following result. 

%%%%%%%%%%%%%%%%
\begin{theorem}% Theorem 1
%%%%%%%%%%%%%%%%
Every collinear and non-zero angular momentum solution of the $n$-body problem
given by a potential that depends only on the mutual distances is homographic.  
\end{theorem}

\noindent{\bf Proof :}
Since ${\bf r}_{i}$ and ${\bf F}_{ij}$ are collinear, 
%----------
\beq
\dot{\bf~ K}_i={\bf M}_i=\sum_{\substack{j=0\\j\neq i}}^n m_i{\bf r}_{i}\times {\bf F}_{ij}=0.
\label{moment}
\eeq
%----------%
Consequently ${\bf K}_i$ is a constant vector. If $K_i$ the component of ${\bf K}_i$ orthogonal to the plane of motion, then $K_i=c_i$, where $c_i$ is a constant. We can also write the component of the angular momentum orthogonal to the plane of motion as
%---------
\beq
K_i=c_i=m_ir_i^2(t)\omega(t),
\label{angmom}
\eeq
%---------%
where the angular velocity $\omega$ is the same for all the bodies, since they 
belong to a line. Let's note that the vectors $r_i$ and the angular velocity $\omega$ may depend on time.

Consider now the ratio between the components of the angular momentum of any two bodies $P_i$ and $P_j$. Using equation (\ref{angmom}), we get
%----------
\beq
{c_i \over c_j}={m_i \over m_j}{r_i^2 \over r_j^2},
\eeq
%----------%
consequently the ratio between distances is
%-----------
\beq
{r_i(t) \over r_j(t)}= \sqrt{m_j c_i \over m_i c_j}.
\eeq
%------------%
Consequently the geometrical configuration of the $n$ bodies remains similar to itself as $t$ varies. This concludes the proof.

\vspace{0.3cm}
The particular Newtonian case of Theorem~1 was proved by Pizzetti in 1904
for the 3-body case, \cite{Pizzetti, Wintner}. Unaware of his result, we found the above shorter, simpler and more general proof. Using Theorem~1, we can now confirm Saari's conjecture in the collinear case for any number $n$ of bodies.

%%%%%%%%%%%%%%%%%%
\begin{corollary}%     COROLLARY 1
%%%%%%%%%%%%%%%%%
Consider the $n$-body problem given by a potential that depends on the
mutual distances alone. Then the only solutions that are collinear, 
have non-zero angular momentum, and constant moment of inertia are the 
relative equilibria. 

\end{corollary}

\noindent{\bf Proof :} The component of the total angular momentum orthogonal to the 
plane of motion can be written as  
%-----------
\beq
K=\sum_{i=0}^N K_i=\omega \sum_{i=0}^N m_ir_i^2 =\omega I=C,
\eeq 
%----------%
where $I$ is the moment of inertia. Since $I$ and $K$ are constant, 
$\omega$ is constant too. The fact that mutual distances are also 
constant, follows immediately from equation (\ref{angmom}). Hence we 
showed that if the moment of inertia is constant, the corresponding 
solutions are relative equilibria. This completes the proof.

\begin{figure}[h]                                                         %
\begin{center}                                                            %
\resizebox{!}{5cm}{\includegraphics{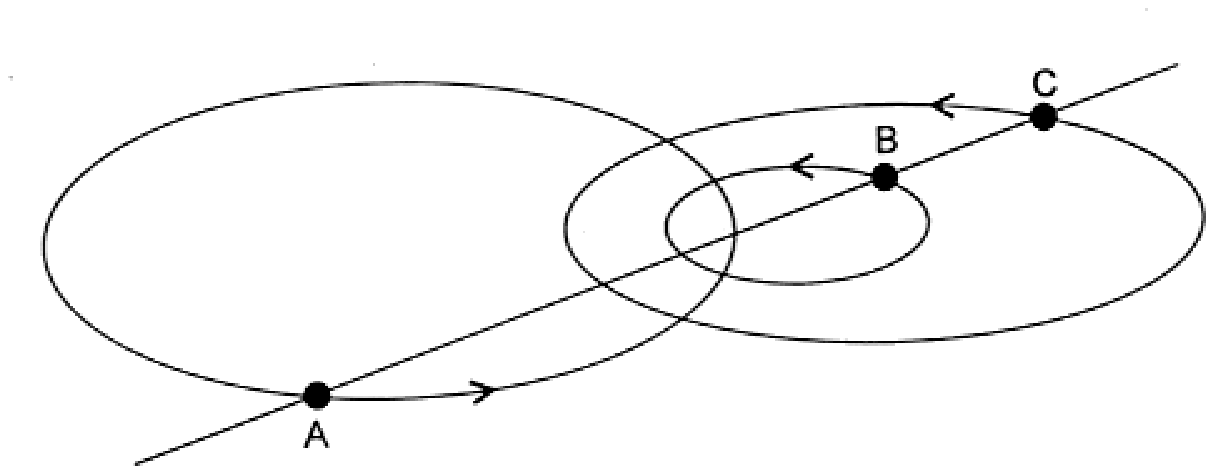}}                            %     FIGURE 
\end{center}                                                              %
\caption[Fig.\ref{Euler}]{The Eulerian Solution of the three-body problem.%
}                                                                         %
\label{Euler}                                                             %
\end{figure}

\vspace{0.3cm}
Corollary 1 completely characterizes the collinear solutions with constant 
moment of inertia, but Theorem 1 is not the strongest result that can be 
proved. Indeed, in 1767 Leonhard Euler showed that if in the Newtonian case 
three bodies of arbitrary masses are arranged initially on a line, as in Figure~\ref{Euler}, such that the ratio $AB/BC$ of their mutual distances 
has a certain value given by a formula depending on the masses, and if suitable 
initial velocities are assigned to the particles, then they will move periodically 
on ellipses remaining on a line at all times. Such motions are called homographic 
motions with central configurations.

%-------------------------------------------------------------------------
\vspace{.3cm}                                                             %
%-------------------------------------------------------------------------
This leads to a natural question: ``Are there any collinear orbits given by homographic solutions with central configurations?'' It turns out that the 
answer is positive if the potential is homogeneous of degree $\alpha \neq -2$. More precisely, we have the following result.
%%%%%%%%%%%%%%%%
\begin{theorem}% Theorem 2
%%%%%%%%%%%%%%%%
The only collinear and non-zero angular momentum solutions of the $n$-body problem 
given by a homogeneous potential of degree $\alpha\neq -2$, are the homographic 
motions with central configurations. In particular if the moment of inertia is 
constant, the solutions are relative equilibria. 
\end{theorem}

{\it Proof:} The first part of the proof follows immediately from Theorem~1 
and Proposition 2.14 in \cite{Albouy}, which shows that if the potential is homogeneous of degree $\alpha\neq -2$, then the homographic solutions are homographic with central configurations. The second part follows from 
Corollary~1. This concludes the proof.

\vspace{0.3cm}
We further present an alternative proof of Theorem~2, which uses Sundman's inequality. Differentiating  the expression of the angular momentum of the body $P_i$ (see equation (\ref{angmom})) with respect to the time variable and using equation (\ref{moment}), we find that
%-------
\beq
\dot K_i=2m_ir_i\dot r_i \omega(t)+m_i r_i^2 \dot\omega(t)=0.
\eeq
%-------%
This implies that
%-------
\beq
\dot r_i=-{r_i \over 2}{\dot \omega \over \omega}.
\label{velocity}
\eeq
%-------%
With these preparations, we can rewrite the kinetic energy,
%-------
\beq
T={1\over 2} \sum_{i=1}^N \left ( m_i \dot r_i^2+m_i r_i^2 \omega^2 \right ),
\eeq
%-------%
in the form
%-------
\beq
T= {C \over 2}\left( {\dot \omega^2 \over 4 \omega^3}+\omega \right),
\label{T}
\eeq
%------%
where we replaced $\dot r_i$ with the expression in (\ref{velocity}) and $I$ with $C/\omega$. 
 
Now consider Sundman's inequality
%-------
\beq
2TI-J^2\geq |C|^2,
\label{sundman}
\eeq
%------%
where
 
%--------
\beq
J= \sum_{i=1}^N m_i~{\bf r}_i\cdot \dot{\bf r}_i=\sum_{i=1}^N m_i r_i \dot r_i
\eeq
%-------%
(see \cite{Albouy, Wintner} for a derivation). Using equation (\ref{velocity}), we can write
%-------
\beq
J=-\sum_{i=1}^N m_ir_i^2 {\dot \omega\over 2\omega}=-I{\dot \omega\over 2\omega}
\eeq
%-------%
and since $C=I\omega$, we get 
%-------
\beq
J=-{\dot\omega \over 2 \omega^2}C.
\label{J}
\eeq
%------%
With the help of equations (\ref{T}) and (\ref{J}), we can write the left hand side of the Sundman's inequality as 
\beq
2TI-J^2={C^2 \over \omega} \left ( {\dot \omega^2 \over 4 \omega^3} +\omega       \right )-\
\left(-{\dot \omega^2 C \over 2 \omega^2} \right)^2= C^2, 
\eeq
which becomes an equality in this case.
Therefore the solutions are homographic (as proved in \cite{Albouy}) 
and since $U$ is homogeneous of degree $\alpha\neq -2$, (\cite{Albouy}, 
Proposition 2.14), they are either solutions with central configurations or 
rigid motions. In the latter case the moment of inertia $I$ is constant and 
we can show as in Corollary~1 that the solutions are relative equilibria. 
This concludes the alternative  proof.

\vspace{.3cm}
Let us remark  that the  homogeneity of
the potential is an essential hypothesis of the theorem
since  there are quasi-homogeneous potentials (see \cite{Diacu2}), in
particular the Lennard-Jones potential, which have
central configurations that depend  on the mutual distances and hence,
in those cases, different and more complex  types of motions might occur.
   
%%%%%%%%%%%%%%%%%%%%%%%%%%%%%%%%%%%%%%%%
\section{A Geometrical Interpretation}
%%%%%%%%%%%%%%%%%%%%%%%%%%%%%%%%%%%%%%%%

In this section we will give a geometrical interpretation of the above results
in the Newtonian case. Our discussions with Don Saari revealed that he had
developed a similar geometric understanding of the problem long before we
found ours.

The moment of inertia can be written in terms of the relative distances as
%--------------------%
\beq
I= {1 \over 2 M} \sum_{i=1}^n\sum_{j=1}^n   m_i m_j r_{ij}^2. 
\eeq
%---------------------
For three collinear bodies, the potential energy and the
moment of inertia contain only three terms, so it is natural to try
to represent the manifolds $I=constant$  and $U=constant$  in the space
of relative  distances. 

%--------------------------------------------------
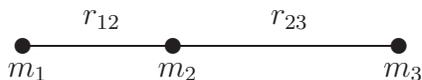
\begin{figure}                                     %
\begin{center}                                     %
\setlength{\unitlength}{1cm}                       %
\begin{picture}(6,2)                               %
\put(0.2,1.5){\line(1,0){5}}                       %
\put(0,1.1){$m_1$}                                 %
\put(1,1.8){$r_{12}$}                              %
\put(0.2,1.5){\circle*{.2}}                        %
\put(3.5,1.8){$r_{23}$}                            %           Figure
\put(5.2,1.5){\circle*{.2}}                        %
\put(5,1.1){$m_3$}                                 %
\put(2.2,1.5){\circle*{.2}}                        %
\put(2,1.1){$m_2$}                                 %
\end{picture}                                      %
\vspace{-0.6cm}                                    %
\caption[Fig.\ref{masses}]{Three bodies on a line  %
for a fixed ordering of the masses.}               %
\label{masses}                                     %
\end{center}                                       %
\end{figure}                                       %
%--------------------------------------------------

\begin{figure}[h]                                                          %
\begin{center}                                                             %
\resizebox{!}{6cm}{\includegraphics{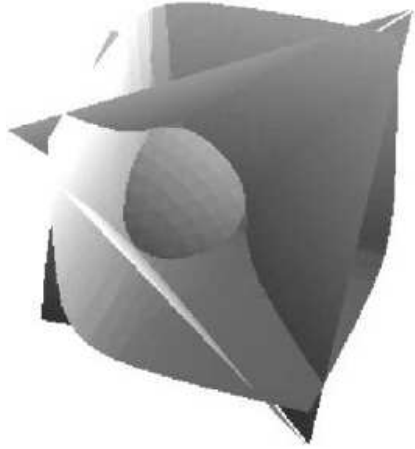}}                            %     FIGURE 
\end{center}                                                               %
\caption[Fig.\ref{manifolds}]{Intersection of the manifolds $U=$constant, $I=$constant  %
with the three planes in the case of equal masses}                         %
\label{manifolds}                                                          %
\end{figure}

Moreover since the  bodies are on  a line, they are subject to additional
constraints. For instance  if one  fixes the ordering of the masses as in 
Figure~\ref{masses}, the condition is that $r_{13}=r_{12}+r_{23}$, which 
means that the  bodies are confined to a plane in the space of relative 
distances. Since three bodies can be ordered in three ways, the corresponding 
conditions define three planes in the space of mutual distances, as we illustrate in Figure \ref{manifolds}, which depicts the intersection of the manifolds $U=constant$, $I=constant$, and the three planes that give the collinear configurations. It is now easy to see that, once ordering the bodies, the collinear relative equilibria correspond to the case in which the intersection 
of $U=constant$ and $I=constant$ restricted to the plane is just one point.
From the purely geometric point of view there are three possible cases
for the outcome of the above intersection: empty, one point, or two points. 
But by Moulton's theorem we know that for each ordering there is only one 
relative equilibrium and thus the relative equilibrium must correspond to 
the case when the intersection of $I=$constant and $U=$constant is tangent 
to the plane (see Figure \ref{manifolds}). This is an alternative proof 
of Saari's Conjecture in the collinear case of three bodies.

%--------------------------------------------------------------------------
\vspace{0.3cm}                                                             %
%--------------------------------------------------------------------------

However, this proof cannot be generalized to $n>3$. Indeed, the geometry becomes more complicated as the number of bodies increases. It is first easy to find the following recurrence relation for the number $M(n)$ of the mutual distances of $n$ bodies:
%-----------------%
\beq
M(n+1)=M(n)+n.
\eeq  
%------------------
Since $M(3)=3$, this recurrence relation can be solved; it leads to the formula
%------------------%
\beq
M(n)={n(n-1)\over 2}.
\eeq
%-------------------
The number of linear relations that must be verified so that the $n$ bodies are on a 
line can also be found. The recurrence relation that gives the number $R(n)$ of linear
relations is
%-------------------%
\beq
R(n+1)=2R(n)-R(n-1) + 1.
\eeq
%-------------------
The initial conditions $R(2)=0$ and $R(3)=1$ lead to the formula
%-----------------%
\beq
R(n)={n(n-1) \over 2} - (n-1).
\eeq
%-----------------
This means that we have $M(n)$ unknowns, $R(n)$ linear relations, and two more 
relations given by $U=constant$ and $I=constant$. Consequently if $n>3$ the 
number of mutual distances is larger then the number of relations. In general
this implies that the problem could have infinitely many solutions, unless the 
intersection is degenerate, in which case the intersection reduces to one point. As we have shown in the previous section, Saari's Conjecture is true in the general case, so the intersection is indeed one point. For $n>3$, however, it is impossible to draw this conclusion only from geometrical considerations.
But it is interesting to remark that the dynamics of the problem leads to 
this unlikely geometric configuration. 
 %%%%%%%%%%%%%%%%%%%%%%%%%%%%%%%
\section*{ Acknowledgments}
%%%%%%%%%%%%%%%%%%%%%%%%%%%%%%%
We are indebted to Don Saari for the many discussions he had with
us about his conjecture in September 2002 while he was a PIMS Distinguished 
Chair at the University of Victoria. Without Don's insight, enthusiasm, and energy, we would have never finalized this research. We would also
like to acknowledge our sponsors: Manuele Santoprete was supported by a 
University of Victoria Fellowship while completing his Ph.D. degree with
Florin Diacu at the University of Victoria, Florin Diacu received partial support from the NSERC of Canada Grant OGP0122045 and from a Research Fellow Grant of the Pacific Institute for the Mathematical Sciences, while Ernesto P\'erez-Chavela was supported in part by Proyecto CONACYT of M\'exico. 

%%%%%%%%%%%%%%%%%%%%%%%%%%%%%%%

\end{document}